\documentclass[10pt,twocolumn,english,aps,prl,superscriptaddress,bibnotes,amsmath,amssymb,floatfix]{revtex4-1}

\usepackage[colorlinks=true,citecolor=blue,linkcolor=magenta]{hyperref}
\usepackage[utf8]{inputenc}
\usepackage[english]{babel}
\usepackage[T1]{fontenc}
\usepackage{amsmath,amsfonts,amssymb}
\usepackage{graphicx}  % for graphics files
\usepackage{tabularx}
\usepackage{lineno}
\usepackage[deletedmarkup=sout, draft]{changes}
\usepackage{siunitx}
\usepackage{upgreek}
\usepackage{moreverb}
\usepackage{mathrsfs}
\usepackage{bm}
\usepackage{makecell}
\usepackage{soul,xcolor}
\usepackage{url}

\begin{document}
\newcommand{\sech}[0]{{\rm sech}}
\def \bwModeN {\Delta \mu}

\title{\textbf{Perspective of high-speed Mach-Zehnder modulators based on nonlinear optics and complex band structures}}

\author{Shuyi Li}
\thanks{These authors contributed equally to this work.}
\affiliation{International Quantum Academy, Shenzhen 518048, China}

\author{Wei Luo}
\thanks{These authors contributed equally to this work.}
\affiliation{Department of Electrical and Electronic Engineering, The Hong Kong Polytechnic University, Hong Kong, China}

\author{Zhenyu Li}
\email[Contact author: ]{ray.shower19@gmail.com}
\affiliation{Linkstar Microtronics Pte. Ltd., Singapore 118222, Singapore}

\author{Junqiu Liu}
\email[Contact author: ]{liujq@iqasz.cn}
\affiliation{International Quantum Academy, Shenzhen 518048, China}
\affiliation{Hefei National Laboratory, University of Science and Technology of China, Hefei 230088, China}

\date{\today}% It is always \today, today,

\begin{abstract}
Optical modulators are essential building blocks for high-capacity optical communication and massively parallel computing. 
Among all types of optical modulators, travelling-wave Mach-Zehnder modulators (TW-MZMs) featuring high speed and efficiency are widely used, and have been developed on a variety of integrated material platforms.
Existing methods to design and simulate TW-MZMs so far strongly rely on the peculiar material properties, and thus inevitably involve complicated electrical-circuit models. 
As a result, these methods diverge significantly.
In addition, they become increasingly inefficient and inaccurate for TW-MZMs with extending length and levitating modulation speed, posing formidable challenges for millimeter-wave and terahertz operation. 
Here, we present an innovative perspective to understand and analyze high-speed TW-MZMs.  
Our perspective leverages nonlinear optics and complex band structures of RF photonic crystals, and is thus entirely electromagnetic-wave-based.
Under this perspective, we showcase the design, optoelectronic simulation and experimental validation of high-speed TW-MZMs based on Si and LiNbO$_3$, and further demonstrate unambiguous advantages in simplicity, accuracy and efficiency over conventional methods.
Our approach can essentially be applied to nearly any integrated material platform, including those based on semiconductors and electro-absorption materials.  
With high-frequency electrode designs and optoelectronic co-simulation, our approach facilitates the synergy and convergence of electronics and photonics, and offers a viable route to constructing future high-speed millimeter-wave and terahertz photonics and quantum systems.
\end{abstract}

\maketitle

%%%%%%%%%%%%%%%%%%%%%%%%%%%%%%%%%%%%%%%%%%%%%%%%%%%%%%%%%%%%%%%%%%%%%%%%%%%%%%%
\section{Introduction}

Optical modulators \cite{XuQ:05, ReedG:10}, which encode information into lightwaves, are pivotal in optical systems and applications such as optical communication \cite{ChengQ:18, WangJ:18}, quantum information \cite{SibsonP:17, WehnerS:18}, and photonic computing \cite{ShenY:17, BogaertsW:20}. 
High-speed modulators can vastly boost signal transmission rates over optical fibers \cite{ZhangM:21, ShiY:22}.
Equally important, for emerging photonics-accelerated artificial intelligence and neural networks \cite{ShastriB:21, FeldmannJ:21, HuangC:22, FarmakidisN:24}, optical modulators play a decisive role in multiplication and nonlinear activation in these systems. 
Innovative photonic computing architectures, such as photonic–electronic deep neural networks \cite{AshtianiF:22}, photonic tensor processing units \cite{DongB:24}, and time-wavelength multiplexing \cite{XuX:21, XuS:22, BaiB:23, ZhuY:24}, utilize modulators to achieve computing speeds far exceeding tera-operations per second (TOPS) and currently approaching peta-operations per second (POPS).
Functioning as dot-product cells in matrix arrays within cross-bar schemes \cite{MoralisM:24, RahimiK:24}, optical modulators have achieved remarkable computational accuracy up to 99\%.
Additionally, high-speed modulators enable ultrafast analogue computing functions such as temporal integration and differentiation. 
These functions endow reconfigurability to multi-purpose microwave photonic processing engines \cite{MarpaungD:19, FengH:24}.

Among all types of optical modulators, travelling-wave Mach-Zehnder modulators (TW-MZMs) are the most widely deployed. 
TW-MZMs offer advantages including reduced thermal sensitivity compared to resonator-based modulators \cite{LiM:20, ZhangW:23}, and simplified RF connection compared to lumped MZMs with segmented electrodes \cite{JacquesM:20, MohammadiA:23}. 
Especially, with progressing wafer-level micro-fabrication technology and emerging nonlinear materials, high-speed TW-MZMs have been developed on various integrated platforms including silicon (Si) \cite{LiM:18, HanC:23}, lithium niobate (LiNbO$_3$) \cite{WangC:18, HeM:19}, lithium tantalate (LiTaO$_3$) \cite{WangC:24, WangC:24b}, barium tantalate (BaTiO$_3$) \cite{EltesF:20, DongZ:23}, III-V semiconductors \cite{SchindlerP:13, OgisoY:20}, electro-optic polymers \cite{LeeM:02, LuG:20} and 2D materials \cite{SunZ:16, SorianelloV:17}.
As such, diverse designs of TW-MZMs have been demonstrated to accommodate and exploit material properties. 
However, these designs differ vastly, particularly in their electrical-circuit models that mimic the materials' response.
For instance, the circuit model for graphene-based TW-MZMs uses the Kubo formula, which relates to the Femi level, temperature, and other factors \cite{SorianelloV:17}. 
In contrast, the circuit model for Si TW-MZMs must account for the nonlinear current-voltage (I-V) relations of PN junctions \cite{ReedG:10}.

Critical to TW-MZM designs is the RF electrode that determines modulation speed or bandwidth.
Successful electrode designs should simultaneously achieve low loss (optical and RF), impedance matching, and group index matching.
Among existing electrode designs, the periodic T-shaped electrodes are often favoured. 
By minimizing RF loss and enabling closer spacing, periodic T-shaped electrodes can overcome the voltage-bandwidth limit in LiNbO$_3$ TW-MZMs \cite{KharelP:21, WangZ:22, ValdezF:23, DuY:24, ZhangY:24}. 
The slow-wave effect facilitates group index matching between RF and optical waves in Si TW-MZMs \cite{DingR:14, PatelD:15, WangX:21, ZhuangD:24}.

Conventionally, designing and modelling periodic T-shaped electrodes have relied on closed-form expressions \cite{ZhuangD:24} or $S$-parameters \cite{DingR:14, PatelD:15, WangX:21, KharelP:21, WangZ:22, ValdezF:23, DuY:24, ZhangY:24}. 
However, both approaches are inefficient or inaccurate. 
The former approach employs conformal mapping to determine electrode capacitance, which becomes increasingly inefficient for complex geometries \cite{YuH:12}. 
Additionally, the expression of resistance and inductance becomes inaccurate at high frequency due to the skin effect \cite{ZhouY:16}. 
The latter approach requires numerical simulation of long electrodes, e.g., 1000 $\mu$m in Ref.~\cite{ZhangY:24}.
This costs excessive computational resources at frequencies exceeding 100 GHz \cite{WangC:18, HeM:19, HanC:23} to 500 GHz \cite{ZhangY:24}.

In parallel, conventional optoelectronic co-simulation is executed to validate electrode designs, which, however, also has limitations. 
The segmented-circuit approach divides optical waveguides and RF electrodes into multiple equivalent circuits with delay intervals,  and simulates them in Verilog-A \cite{ZhuK:15, LinS:17} or SPICE \cite{MingD:24}. 
This approach is inaccurate due to the different physics between I-V and wave signals.
The transmission-line-circuit (TLC) approach requires numerically solving the electrode's I-V equations combined with PN junctions \cite{BahramiH:16, ZhangQ:19},  and thus is inefficient.
In both approaches,  computationally heavy convolution is necessitated to account for the electrode's dispersion.

Here, we present an innovative perspective to understand and analyze high-speed TW-MZMs. 
Our perspective overcomes the above-mentioned inaccuracy and inefficiency in the design and simulation of TW-MZMs, and can be universally applied to any integrated material platforms.
Our perspective is rooted in nonlinear optics \cite{HerrT:12, MossD:13} and complex band structures (CBS) in photonic crystals \cite{SukhoivanovI:09, ChebenP:18}, both of which are derived from Maxwell’s equations -- the ``first principles'' of electromagnetic waves.
We view the modulators as \textit{nonlinear RF photonic-crystal waveguides} – a paradigm shift from conventional electrical-circuit models to a fully electromagnetic-wave model. 
With this perspective, we showcase a unified design and simulation process for Si and LiNbO$_3$ TW-MZMs.
We develop a CBS-based simulation method and demonstrate a 100-times simulation speedup on LiNbO$_3$ TW-MZMs of 500 GHz bandwidth. 
We further establish a nonlinear-optics-based optoelectronic co-simulation that obviates computationally heavy I-V equations for PN junctions and convolution calculations.
Finally, we present a co-simulated eye diagram for bidirectional RF and optical waves and experimental validation.

%%%%%%%%%%%%%%%%%%%%%%%%%%%%%%%%%%%%%%
\section{Principles of travelling-wave Mach-Zehnder modulators}

%%%%%%%%%%%%%%%%%%%%%%%%%%%%%%%%
\begin{figure*}[t!]
\centering
\includegraphics[width=15.5cm]{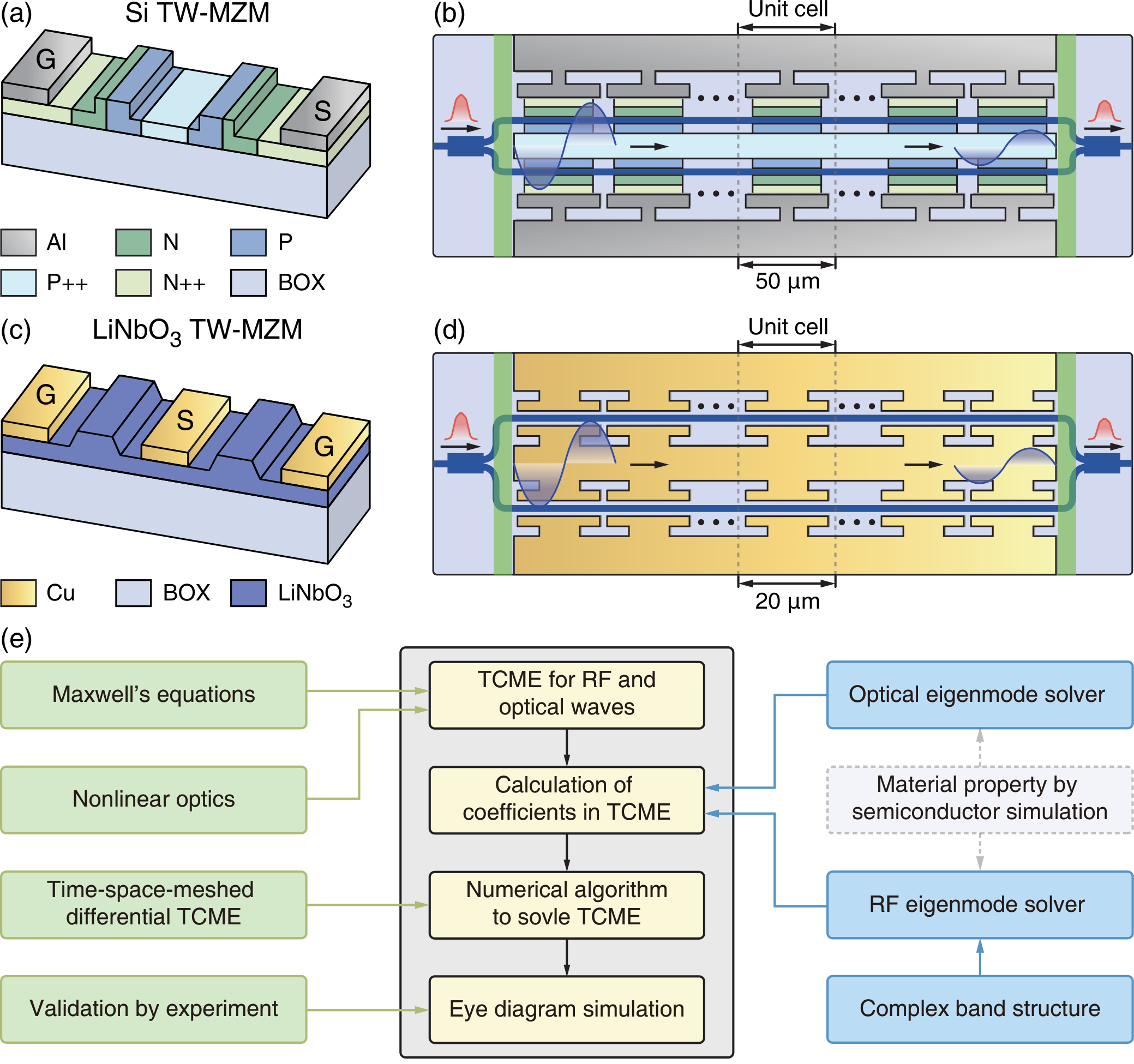} 
\caption{
 Schematics, layouts and our design process of Si and LiNbO$_3$ TW-MZMs. 
(a, c) Cross-sections of the Si TW-MZM with GS RF electrodes (a), and the x-cut LiNbO$_3$ TW-MZM with GSG electrodes (c).
(b,d) Schematics and layouts of the Si (b) and LiNbO$_3$ (d) TW-MZMs, including their periodic T-shaped RF electrodes and optical waveguides. 
The electrode's unit cell is outlined within dashed boxes, and the green regions represent the RF wave's input ports and absorbers at output.
(e) Our process flowchart to design and simulate high-speed TW-MZMs.
}
\label{fig1}
\end{figure*}
%%%%%%%%%%%%%%%%%%%%%%%%%%%%%%%%

Figure \ref{fig1}(a,c) presents the cross-sections of a Si TW-MZM with ground-signal (GS) RF electrode, and an x-cut LiNbO$_3$ TW-MZM with ground-signal-ground (GSG) RF electrodes.
Figure \ref{fig1}(b,d) depicts the schematics and layouts of the Si and LiNbO$_3$ TW-MZMs, including their periodic T-shaped RF electrodes (formed by an array of unit cells) and optical waveguides. 
The Mach-Zehnder interferometer (MZI) consists of two multimode interferometers (MMI) and two optical waveguide arms.
Amplitude modulation of optical signal is achieved via tuning the phase difference between the MZI's two arms. 
When voltage is applied between the G and S electrodes, phase difference is induced due to refractive index change of optical waveguides. 
For Si TW-MZMs, an array of PN junctions in the lateral push-pull configuration is sandwiched by the GS electrodes.
When alternating voltage is applied on the GS electrodes, the electron and hole's distribution in the PN junctions alters accordingly,  which changes the refractive index of the Si waveguides due to the plasma dispersion effect. 
For LiNbO$_3$ TW-MZMs, the refractive index of LiNbO$_3$ waveguides is varied due to the Pockels effect.

Here, we interpret the operational principle of Si and LiNbO$_3$ TW-MZMs as following.
The RF electrodes are treated as \textit{periodic sub-wavelength grating waveguides} \cite{ChebenP:18}, where the RF wave propagates.
Thus the dispersion of RF waveguides (electrodes) is naturally addressed by deriving the RF wave's group velocity in the waveguide. 
This treatment avoids conventionally required, complicated convolution calculation \cite{ZhuK:15, LinS:17, MingD:24, BahramiH:16, ZhangQ:19}. 
For Si TW-MZMs, the array of PN junctions is treated as an \textit{equivalent, fourth-order nonlinear RF material} (i.e., with permittivity and conductivity), by replacing the nonlinear I-V equations of PN junctions with self- and cross-phase modulation (SPM and XPM) of the RF pump.
In this perspective, optical modulation in TW-MZMs is a direct consequence of nonlinear interaction between the RF (as the pump) and optical waves (as the signal). 
Therefore, a universal theoretical framework based on nonlinear optics is capable of modelling TW-MZMs on nearly any material platform besides Si and LiNbO$_3$.  

In this framework, the TW-MZM design process involves four steps as illustrated in Fig.\ref{fig1}(e) grey box. 
First, we use temporal coupled-mode equations (TCME) -- derived from Maxwell’s equations -- to describe nonlinear interaction between RF and optical waves. 
Second, coefficients in TCME are determined using optical and RF eigenmode solvers.
For Si TW-MZMs, the eigenmode solvers require extra semiconductor simulation.
Third, a numerical algorithm solves the differential-form TCME in a time-space mesh grid. 
Finally, with a digital input signal, the simulation results in the eye diagram. 
In the following sections, we illustrate each step in details.

%%%%%%%%%%%%%%%%%%%%%%%%%%%%%%%%%%%%%%%%%%%%%%%%%%%%%%%%%%%%%%%%%%%%%%%%%%%%%%%
\section{Temporal coupled-mode equations}

We use temporal coupled-mode equations (TCME) to describe the nonlinear interaction between RF and optical waves, which are both electromagnetic waves but with vastly different frequencies.
Table \ref{Tab1} summarizes the definition of each symbol used to derive TCME.
In a TW-MZM, we consider one static electric field $E_\text{dc}$ induced by the bias voltage $V_\text{dc}$, 
one optical wave $E_\text{opt}$ with amplitude $A_\text{opt}$, 
and one forward RF wave with amplitude $A_\text{rf,F}$.
The RF and optical waves experience attenuation ($\alpha_\text{rf}$ and $\alpha_\text{opt}$) during propagation. 
Caused by impedance mismatch, the forward RF wave is accompanied by a reflected RF wave with amplitude $A_\text{rf,R}$. 
The forward and reflected RF waves share the same electric field $E_\text{rf}$.  
Note that $E_\text{dc}$, $E_\text{opt}$, and $E_\text{rf}$ are all dominant components of the normalized electric field distribution.
The TCME for $A_\text{opt}$, $A_\text{rf,F}$ and $A_\text{rf,R}$ are

\begin{subequations}
\begin{align}
\left( \frac{\partial}{\partial z} + \alpha_\text{rf} + \frac{1}{v_\text{g,rf}} \frac{\partial}{\partial t} \right) A_\text{rf,F}=&P_\text{rf,F}^{(2)}+P_\text{rf,F}^{(3)}+P_\text{rf,F}^{(4)}~, \label{Eq.RFF_TCME} \\
\left(-\frac{\partial}{\partial z} + \alpha_\text{rf} + \frac{1}{v_\text{g,rf}} \frac{\partial}{\partial t} \right) A_\text{rf,R}=&P_\text{rf,R}^{(2)} + P_\text{rf,R}^{(3)}+P_\text{rf,R}^{(4)}~, \label{Eq.RFR_TCME} \\
\left( \frac{\partial}{\partial z} + \alpha_\text{opt} +\frac{1}{v_\text{g,opt}} \frac{\partial}{\partial t} \right)A_\text{opt}=&P_\text{opt}^{(2)}+P_\text{opt}^{(3)}~. \label{Eq.Opt_TCME}
\end{align}
\end{subequations}
Here, we designate that RF and optical waves propagate in the $z$-direction, thus $A_\text{opt}$, $A_\text{rf,F}$, and $A_\text{rf,R}$ are $z$-dependent. 
The derivation of Eq. (\ref{Eq.RFF_TCME}--\ref{Eq.Opt_TCME}) is found in Supplementary Note 1.
The reason why the fourth-order nonlinear term $P_\text{opt}^{(4)}$ is neglected in Eq. (\ref{Eq.Opt_TCME}) is illustrated later with Eq. (\ref{Eq.neff_opt}).

%%%%%%%%%%%%%%%%%%%%%%%%%%%%%%%%%
\begin{table*}[t!]
\centering
\caption{Summary of symbols used in TCME}
\begin{ruledtabular}
\begin{tabular}{cccc}
Symbol&Definition&Symbol&Definition\\
\hline
\makecell{$A_\text{rf,F}$, $A_\text{rf,R}$, $A_\text{opt}$} & \makecell{Forward RF / reflected RF / optical\\wave's amplitude.} &
     \makecell{$P_\text{rf,F}^{\text{(m)}}$, $P_\text{rf,R}^{\text{(m)}}$,  $P_\text{opt}^{\text{(m)}}$} &  \makecell{Forward RF / reflected RF / optical \\wave's $\text{m}^\text{th}$-order nonlinear term.}\\
    $E_\text{rf}, E_\text{opt}$ & \makecell{RF / optical wave's electric field.} & $V_\text{rf,F}, V_\text{rf,R}$ & \makecell{Forward / reflected RF wave's voltage.}\\
    $E_\text{dc}$ & \makecell{Static electric field.} & $V_\text{dc}$ & Bias voltage.\\
    $\omega_\text{rf}, \omega_\text{opt}$ & \makecell{RF / optical wave's  angular frequency.} & $k_\text{0,rf}, k_\text{0,opt}$ & \makecell{RF / optical wave's wave-number.}\\
    $\alpha_\text{rf}, \alpha_\text{opt}$ & \makecell{RF / optical wave's\ attenuation.} & $\beta_\text{rf}, \beta_\text{opt}$ & \makecell{RF / optical wave's phase constant.}\\
    $v_\text{g,rf}, v_\text{g,opt}$ & \makecell{RF / optical wave's group velocity.} & $n_\text{g,rf}, n_\text{g,opt}$ & \makecell{RF / optical wave's group index.}\\
    $n_\text{eff,rf}, n_\text{eff,opt}$ & \makecell{RF / optical wave's effective index.} & $n_\text{r,rf}, n_\text{r,opt}$ & \makecell{RF / optical wave's phase index.}\\
    $Z_\text{c}$ & RF impedance. & $\chi^{(m)}$ & \makecell{$\text{m}^\text{th}$-order nonlinear susceptibility.} \\ 
\end{tabular}
\end{ruledtabular}
\label{Tab1}
\end{table*}
%%%%%%%%%%%%%%%%%%%%%%%%%%%%%%%%

Next, we obtain the $P_\text{rf,F}^{\text{(m)}}$ (where $m=2,3, 4$) terms in Eq. (\ref{Eq.RFF_TCME}). 
Since RF frequency mixing is negligible in TW-MZMs, we only consider SPM and XPM for $A_\text{rf,F}$. 
Meanwhile, small RF impedance mismatch (i.e. $Z_\text{c}\approx50$ $\Omega$) can be easily achieved in practice, leading to $A_\text{rf,R}<A_\text{rf,F}$.
Therefore, XPM on $A_\text{rf,F}$ due to $A_\text{rf,R}$ can be ignored.  
Meanwhile,  moderate optical power in TW-MZMs is insufficient to induce nonlinear optical processes such as four-wave mixing and harmonic generation, thus $A_\text{rf,F}$ is not affected by $A_\text{opt}$.  
Overall, the $P_\text{rf,F}^{\text{(m)}}$ terms including the bias caused by $V_\text{dc}$ and SPM of $A_\text{rf,F}$ can be expressed as

\begin{equation}
\left\{
\begin{aligned}
P_\text{rf,F}^{(2)}=&2i\Lambda_{\text{dr,12}}V_\text{dc}A_\text{rf,F}~,\\
P_\text{rf,F}^{(3)}=&3i\Lambda_{\text{r,4}}\left|A_\text{rf,F}\right|^2A_\text{rf,F}+3i\Lambda_\text{dr,22}V_\text{dc}^2A_\text{rf,F}~,\\
P_\text{rf,F}^{(4)}=&12i\Lambda_{\text{dr,14}}V_\text{dc}\left|A_\text{rf,F}\right|^2A_\text{rf,F}+4i\Lambda_\text{dr,32}V_\text{dc}^3A_\text{rf,F}~,\\
\end{aligned}
\right.
\label{Eq.PrfF}
\end{equation}
where $i$ is the imaginary unit. 
The $\Lambda$ coefficients are defined as

\begin{equation} 
\label{chi_def}
\Lambda_\text{drp,hjk} \equiv  \iint {\rm dxdy} \, \varepsilon_0\chi^{\left(\text{h+j+k-1}\right)}E_\text{dc}^\text{h}E_\text{rf}^\text{j}E_\text{opt}^\text{k}~.
\end{equation}
The integration is over the TW-MZM's cross-section ($xy$-plane), and the exponents $(\text{h, j, k})\in\mathbb{N}$. 
The subscript $(\text{d, r, p})$ marks whether $E_\text{dc}$ (labeled with d), $E_\text{rf}$ (labeled with r) and $E_\text{opt}$ (labeled with p) are involved. % in Eq. (\ref{chi_def}). 
We also stipulate that if any of $(\text{h, j, k})$ is zero, the corresponding indices $(\text{d, r, p})$ are omitted.
For example, 

\begin{equation}
\left\{
\begin{aligned}
\Lambda_\text{dp,12}=&\iint {\rm dxdy} \, \varepsilon_0\chi^{\left(\text{2}\right)}E_\text{dc}E_\text{opt}^\text{2}~,\\
\Lambda_\text{r,4}=&\iint {\rm dxdy} \, \varepsilon_0\chi^{\left(\text{3}\right)}E_\text{rf}^\text{4}~.
\end{aligned}
\right.
\end{equation}
Additionally, $(\text{h}+\text{j}+\text{k}-1)=2,\text{ }3,\text{ and }4$ correspond to $\chi^{(2)}$, $\chi^{(3)}$ and $\chi^{(4)}$ processes, respectively. %modify SYL

%%%%%%%%%
The $P_\text{rf,R}^{\text{(m)}}$ terms in Eq. (\ref{Eq.RFR_TCME}) can be expressed as

\begin{equation}
\left\{
\begin{aligned}
P_\text{rf,R}^{(2)}=&2i\Lambda_{\text{dr,12}}V_\text{dc}A_\text{rf,R}~,\\
P_\text{rf,R}^{(3)}=&3i\Lambda_\text{r,4}\left|A_\text{rf,R}\right|^2A_\text{rf,R}+3i\Lambda_\text{dr,22}V_\text{dc}^2A_\text{rf,R}\\
&+6i\Lambda_\text{r,4}\left|A_\text{rf,F}\right|^2A_\text{rf,R}~,\\
P_\text{rf,R}^{(4)}=&12i\Lambda_\text{dr,14}V_\text{dc}\left|A_\text{rf,R}\right|^2A_\text{rf,R}+4i\Lambda_\text{dr,32}V_\text{dc}^3A_\text{rf,R}\\
&+24i\Lambda_\text{dr,14}V_\text{dc}\left|A_\text{rf,F}\right|^2A_\text{rf,R}~.\\
\end{aligned}
\right.
\end{equation}
Here we keep the terms $6i\Lambda_\text{r,4}\left|A_\text{rf,F}\right|^2A_\text{rf,R}$ and $24i\Lambda_\text{dr,14}V_\text{dc}\left|A_\text{rf,F}\right|^2A_\text{rf,R}$ that describe XPM on $A_\text{rf,R}$ due to $A_\text{rf,F}$. 

%%%%%%%%%%%%
The $P_\text{opt}^{\text{(m)}}$ terms in Eq. (\ref{Eq.Opt_TCME}) can be expressed as

\begin{equation}
\left\{
\begin{aligned}
P_\text{opt}^{(2)}=&2i\Lambda_\text{dp,12}V_\text{dc}A_\text{opt}+2i\Lambda_\text{rp,12}\left(V_\text{rf,F}+V_\text{rf,R}\right)A_\text{opt}~,\\
P_\text{opt}^{(3)}=&
3i\Lambda_\text{dp,22}V_\text{dc}^2A_\text{opt}+3i\Lambda_\text{rp,22}\left(V_\text{rf,F}+V_\text{rf,R} \right)^2A_\text{opt}\\
&+6i\Lambda_\text{rdp,112}V_\text{dc}\left(V_\text{rf,F}+V_\text{rf,R}\right)A_\text{opt}~,
\end{aligned}
\right.
\label{Eq.Popt}
\end{equation}
where the voltage $V_\text{rf,F}$ / $V_\text{rf,R}$ of $A_\text{rf,F}$ / $A_\text{rf,R}$ along the RF waveguide (electrode) is derived as

\begin{equation}
\label{V_def}
\left\{
\begin{aligned}
V_\text{rf,F}\left(z,t\right)=&A_\text{rf,F}\left(z,t\right)\exp{\left(i\beta_\text{rf}z-i\omega_\text{rf}t\right)}+\text{c.c.}\\
V_\text{rf,R}\left(z,t\right)=&A_\text{rf,R}\left(z,t\right)\exp{\left(-i\beta_\text{rf} z-i\omega_\text{rf}t\right)}+\text{c.c.}\\
\end{aligned}
\right.
\end{equation}
In Eq. (\ref{Eq.Popt}), all nonlinear terms relating $A_\text{rf,F}$, $A_\text{rf,R}$ and $V_\text{dc}$ with $A_\text{opt}$ are kept.  
This is to account for phase mismatch between the RF and optical waves, since the typical TW-MZM length of a few millimeters is comparable to the RF wavelength. 
Meanwhile, the optical wave's SPM is ignored due to insufficient optical power.

We note that the computation of $\Lambda_\text{drp,hjk}$ via Eq. (\ref{chi_def}) is difficult, besides that the value of PN junctions' $\chi$ is unknown. 
Instead, the values of $\Lambda_\text{drp,hjk}$ can be obtained as follows. 
First,  discrete values of effective index change $\Delta n_\text{eff,rf}$ and $\Delta n_\text{eff,opt}$ with different $V_\text{dc}$ values are numerically calculated using optical and RF eigenmode solvers. 
Then polynomial fits of  $\Delta n_\text{eff,rf}$ and $\Delta n_\text{eff,opt}$ with $V_\text{dc}$ are performed, as

\begin{subequations} 
\begin{align}
\Delta n_\text{eff,opt}=&\frac{1}{k_\text{0,opt}}\left(2\Lambda_\text{dp,12}V_\text{dc}+3\Lambda_\text{dp,22}V_\text{dc}^2\right)~,\label{Eq.neff_opt}\\
\Delta n_\text{eff,rf}=&\frac{1}{k_\text{0,rf}}\left(2\Lambda_\text{dr,12}V_\text{dc}+3\Lambda_\text{dr,22}V_\text{dc}^2+4\Lambda_\text{dr,32}V_\text{dc}^3\right)~. \label{Eq.neff_RF}
\end{align}
\end{subequations}
The derivation is found in Supplementary Note 1. 
In details, $\Delta n_\text{eff,opt}$ versus $V_\text{dc}$ is numerically calculated using an optical eigenmode solver with the Soref and Bennett’s equations for Si \cite{SorefR:1987}, or with the Pockels effect for LiNbO$_3$ \cite{WeisR:1985}, as illustrated in Supplementary Note 2 and 3.
A second-order polynomial fit of $\Delta n_\text{eff,opt}$ versus $V_\text{dc}$ yields the values of $\Lambda_\text{dp,12}$ and $\Lambda_\text{dp,22}$. 
Based on the simulation result in Supplement Materials Note 3, the second-order polynomial fit is sufficient, justifying the omission of $P_\text{opt}^{(4)}$.
We further assume $\Lambda_\text{dp,12}=\Lambda_\text{rp,12}$ and $\Lambda_\text{dp,22}=\Lambda_\text{rp,22}=\Lambda_\text{drp,112}$.

The CBS-based RF eigenmode solver to calculate $\Delta n_\text{eff,rf}$ versus $V_\text{dc}$ is illustrated in the next section. 
Similarly, a third-order polynomial fit of $\Delta n_\text{eff,rf}$ versus $V_\text{dc}$ yields the values of $\Lambda_\text{dr,12}$, $\Lambda_\text{dr,22}$ and $\Lambda_\text{dr,32}$. 
Considering that RF waves have quasi-TEM field distribution \cite{PozarD:21}, and neglecting dispersion of $\chi$ from DC to RF, we have $\Lambda_\text{r,4} = \Lambda_\text{dr,22}$ and $\Lambda_\text{dr,14} = \Lambda_\text{dr,32}$. 

%%%%%%%%%%%%%%%%%%%%%%%%%%%%%%%%%%%%%%%%%%%%%%%%%%%%%%%%%%%%%%%%%%%%%%%%%%%%%%%
\section{Complex band structures of periodic T-shaped RF electrodes}
%%%%%%%%%%%%%%%%%%%%%%%%%%%%%%%%%%%%%
\begin{figure*}[t!]
\centering
\includegraphics[width=15.5cm]{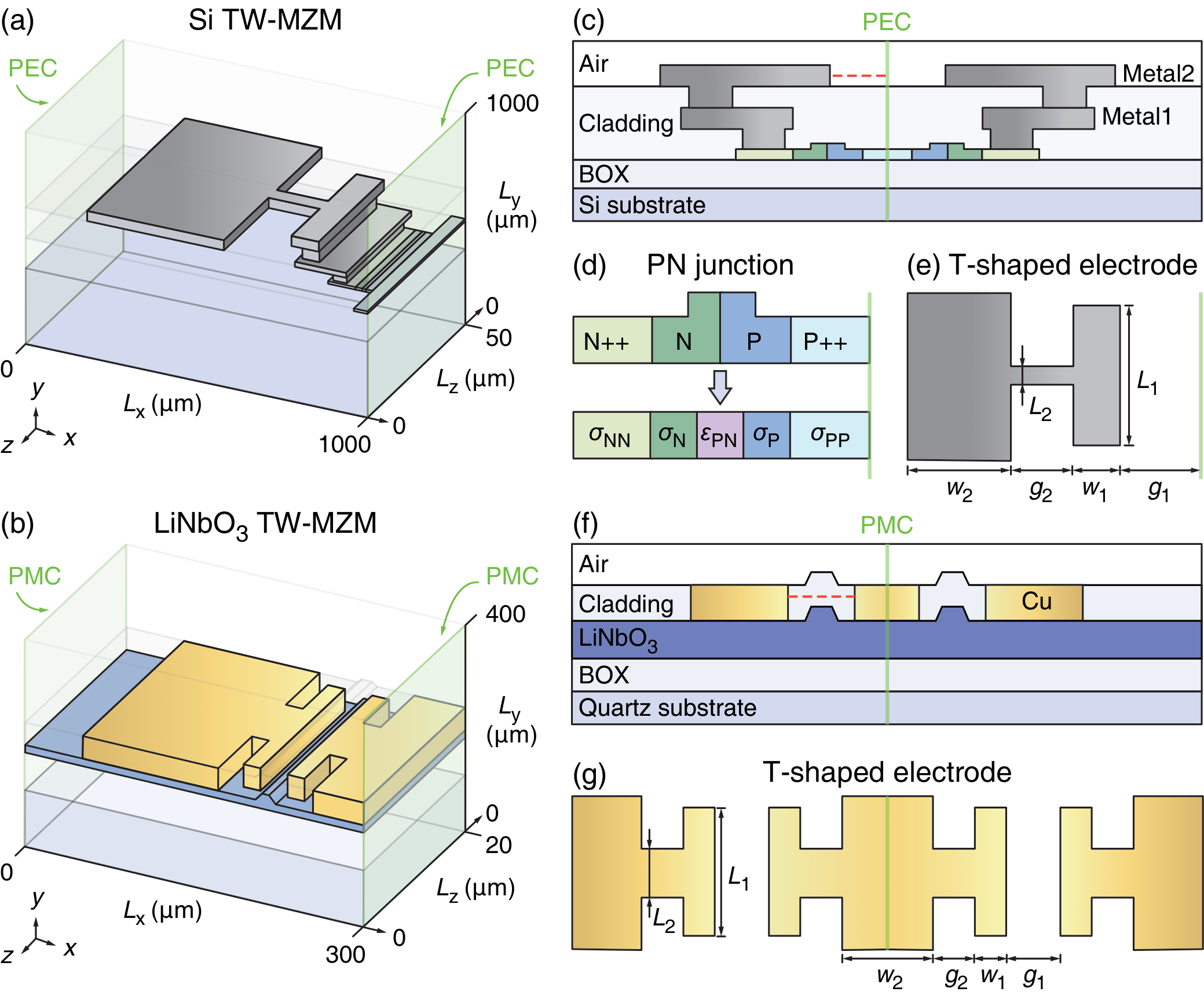} 
\caption{
Simulation models of periodic T-shaped RF electrodes.  
(a,b) RF simulation models for Si (a) and LiNbO$_3$ (b) TW-MZMs with half of the electrode's unit cell.
Green planes are PEC (a) or PMC (b) boundary condition. 
Other boundary planes are set as periodic condition. 
(c,f) Cross-section and layer structure of Si (c) and LiNbO$_3$ (f) TW-MZMs.
(d) The transformation of the PN junction to an equivalent RF material with permittivity and conductivity.
(e, g) Layout and dimension of the unit cell of periodic T-shaped electrodes for Si (GS, c) and LiNbO$_3$ (GSG, f) TW-MZMs.
}
\label{fig2}
\end{figure*}
%%%%%%%%%%%%%%%%%%%%%%%%%%%%%%%%%%%%%

%%%%%%%%%%%%%%%%%%%%%%%%%%%%%%%%%%%%
\begin{figure*}[t!]
\centering
\includegraphics[width=15.5cm]{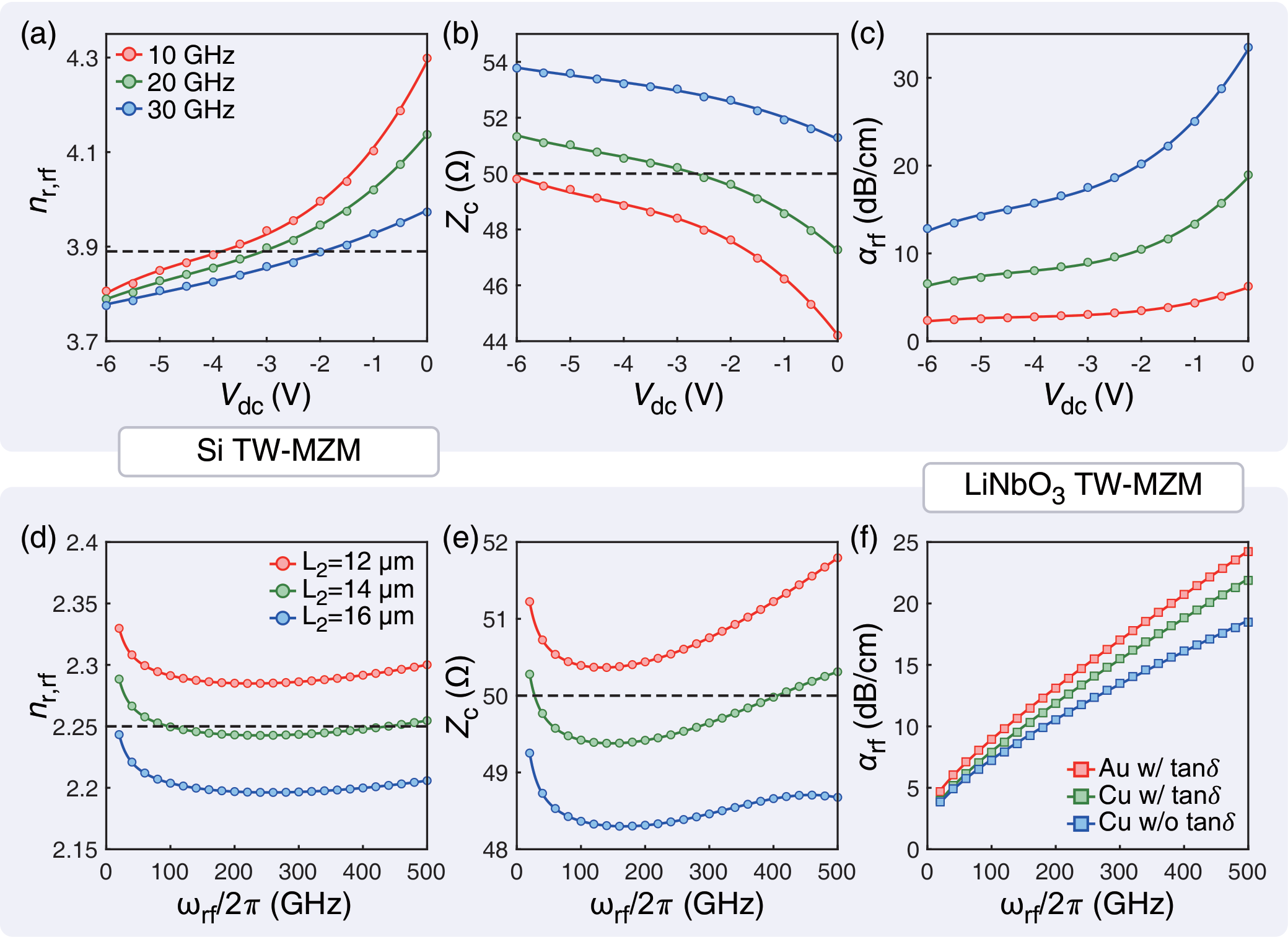} 
\caption{ 
Simulation results of periodic T-shaped RF electrodes.  
(a-c) Calculated $n_\text{r,rf}$, $Z_\text{c}$ and $\alpha_\text{rf}$ with varying $V_\text{dc}$ and $\omega_\text{rf}/2\pi=10, 20, 30$ GHz, for Si TW-MZMs. 
(d, e) Calculated $n_\text{r,rf}$ and $Z_\text{c}$ with varying $\omega_\text{rf}/2\pi\in[20, 500]$ GHz and $L_2=12, 14, 16$ $\mu$m, for LiNbO$_3$ TW-MZMs.
(f) With $L_2=14$ $\mu$m and $\omega_\text{rf}/2\pi\in[20, 500]$ GHz, the calculated $\alpha_\text{rf}$ of Au with tan$\delta$, Cu with tan$\delta$, and Cu without tan$\delta$.
}
\label{fig3}
\end{figure*}
%%%%%%%%%%%%%%%%%%%%%%%%%%%%%%%%%%%%

High-speed TW-MZMs require minimal attenuation $\alpha_\text{rf}$, matched impedance $Z_\text{c}=50$ $\Omega$, and matched indices $n_\text{r,rf}=n_\text{g,opt}$ (with $n_\text{g,opt}=3.89$ for Si and $2.25$ for LiNbO$_3$) \cite{GhioneG:09}.
In this section, we present an RF eigenmode solver to numerically compute these parameters. 
The eigenmodes of Maxwell’s equations are computed under Bloch’s condition \cite{ZhangZ:1990}. 
The solutions are eigen-vectors varying with RF frequency, and thus form complex band structures (CBS).
With the CBS and field distribution of the eigenmodes (also known as Bloch modes), the values of $n_\text{eff,rf}$, $n_\text{r,rf}=\text{Re}(n_\text{eff,rf})$, $\alpha_\text{rf}$, $Z_\text{c}$, $n_\text{g,rf}$ and $\beta_\text{rf}$ used in TCME can be acquired.

The CBS is calculated through finite-element simulation (FEM) of weak-form Maxwell’s equations \cite{FietzC:11}.
First, to obtain the weak-form Maxwell’s equations, we derive the RF vector wave equation as 

\begin{equation}
\nabla \times \left( \nabla\times \boldsymbol{E} \right)-k_\text{0,rf}^2\varepsilon_\text{re} \boldsymbol{E}=0~,
\label{Eq.v_Maxwell}
\end{equation}
where $\varepsilon_\text{re}=\varepsilon_\text{r}-\sigma/(i\omega_\text{rf}\varepsilon_0)$ is the effective relative permittivity, $\varepsilon_0$ is the vacuum permittivity, $\varepsilon_\text{r}$ is the relative permittivity, and $\sigma$ is the conductivity. 
Using Bloch’s theorem, the RF electric field is expressed as

\begin{equation}
\boldsymbol{E}=\boldsymbol{u}\exp{\left(ik_\text{z} z\right)}~,
\end{equation}
where $\boldsymbol{u}$ is the Bloch mode's field distribution, 
and $k_\text{z}$ is the eigen-vector's $z$-component. 
Volume integration over the entire simulation region (described later) yields the weak-form Maxwell’s equation, as

\begin{equation}
\begin{split}
\label{Eq.weak_form}
&\iiint {\rm d}V \, \left((\nabla-ik_\text{z}\boldsymbol{\hat{z}})\times \boldsymbol{w}\right)\cdot\left(\left(\nabla+ik_\text{z}\boldsymbol{\hat{z}}\right)\times \boldsymbol{u}\right)\\
&-k_\text{0,rf}^2\boldsymbol{w}\cdot\varepsilon_\text{re} \boldsymbol{u}=0~,
\end{split}
\end{equation}
where $\boldsymbol{w}$ is the finite-element test function of $\boldsymbol{u}$ \cite{FietzC:11}.
The derivation is found in Supplementary Note 4. 
The $k_\text{z}$ is calculated via FEM of Eq. \ref{Eq.weak_form} in COMSOL Multiphysics \cite{FietzC:11, LiS:20}.
Once the electric field is determined, the magnetic field is derived according to Maxwell’s equations.

Figure \ref{fig2}(a, b) shows the RF simulation models for the Si and LiNbO$_3$ TW-MZMs. 
Due to mirror symmetry, the models contain only half of the electrode's unit cell,  to save computation resources.
Perfect electric / magnetic conductor (PEC / PMC) condition, marked as green planes in Fig. \ref{fig2}(a, b), is used for mirror symmetry of the GS / GSG electrode. 
The region dimension is $(L_\text{x}, L_\text{y}, L_\text{z})$=$(1000, 1000, 50)$ $\mu$m in Fig. \ref{fig2}(a), and $(300, 400, 20)$ $\mu$m in Fig. \ref{fig2}(b).

Figure \ref{fig2}(c) shows the simulated cross-section of the Si TW-MZM, consisting of two push-pull PN junctions, Metal 1 and 2 (aluminium, Al), and two via holes for electrical connection through the SiO$_2$ cladding.  
Beneath the PN junctions is buried SiO$_2$ (BOX) on a Si substrate.
Figure \ref{fig2}(d) illustrates the transformation of the PN junction to an equivalent RF material (see Supplementary Note 2),  which spares  computationally heavy, 3D semiconductor simulation.
The electrode's dimension is depicted in Fig. \ref{fig2}(e), with $(L_1, L_2, g_1, g_2, w_1, w_2)=(47, 3,1, 22, 14.5, 200)$ $\mu$m.
Figure \ref{fig2}(f) shows the simulated cross-section of the LiNbO$_3$ TW-MZM with copper (Cu) electrodes. %quartz substrate 
We use loss tangent values of tan$\delta=0.006$ for SiO$_2$ and 0.008 for LiNbO$_3$, following Ref.~\cite{ZhangY:22}.
Figure \ref{fig2}(g) presents the T-shaped GSG electrodes.
The S and G electrodes share the same dimension $(L_1, L_2, g_1, g_2, w_1, w_2)=(18, 14, 5.5, 3, 2, 25)$ $\mu$m.
The ridge / slab thickness is 220 / 90 nm for PN junctions, and 500 / 250 nm for LiNbO$_3$. 
Both the PN junctions and LiNbO$_3$ are treated as thin-layer boundary condition (see Supplementary Note 4).

With simulated $k_\text{z}$, we have $\beta_\text{rf}=\text{Re}(k_\text{z})$, 
$\alpha_\text{rf}=\text{Im}(k_\text{z})$, 
$n_\text{eff,rf}=k_\text{z}/k_\text{0,rf}$, 
$n_\text{r,rf}=\text{Re}(n_\text{eff,rf})$,
$n_\text{g,rf}={\rm c} \beta_\text{1,rf}$,
where $ \beta_\text{1,rf}$ is the first-order Taylor expansion of $\beta_\text{rf}(\omega_\text{rf})$.
Thus, the assumption of $n_\text{g,rf} \approx n_\text{r,rf}$ in Ref. ~\cite{PatelD:15, LiM:18} is inaccurate.
Then $Z_\text{c} = V^2 / P_\text{z}$ is calculated,  where

\begin{subequations}
\begin{align}
P_\text{z}&=2\iint_S {\rm d}x{\rm d}y S_\text{z} \label{Eq.Pz_def} \\
V&= \left\{
\begin{aligned}
\int_C {\rm d}x E_\text{x}, ~~&\text{with GSG electrodes} \\[2mm]
2 \int_C {\rm d}x E_\text{x}, ~~&\text{with GS electrodes}
\end{aligned}
\right.
\label{Eq.V_def}
\end{align}
\end{subequations}
The surface $S$ in Eq. (\ref{Eq.Pz_def}) covers the simulation region's cross-section at $L_\text{z}=0$ $\mu$m, and $S_\text{z}$ is the $z$-component of the Poynting vector. 
Since $V$ is path-independent for a quasi-TEM field, the path curve $C$ in Eq. (\ref{Eq.V_def}) is chosen as the red dashed lines in Fig. \ref{fig2}(a, b).

For Si TW-MZMs, Fig. \ref{fig3}(a--c) presents the simulation results of $n_\text{r,rf}$, $Z_\text{c}$ and $\alpha_\text{rf}$, with varying $V_\text{dc}$ and $\omega_\text{rf}/2\pi=10, 20, 30$ GHz.
The conditions $n_\text{r,rf}=n_\text{g,opt}=3.89$ and $Z_\text{c}=50$ $\Omega$ are satisfied at $V_\text{dc}=-3$ V and $\omega_\text{rf}/2\pi=20$ GHz.
Parameter sweep of $\omega_\text{rf}/2\pi\in[10, 120]$ GHz and $V_\text{dc}\in[-6, 0]$ V is found in Supplementary Note 4.

For LiNbO$_3$ TW-MZMs, Fig. \ref{fig3}(d, e) presents the simulation results of $n_\text{r,rf}$ and $Z_\text{c}$ with varying $\omega_\text{rf}/2\pi\in[20, 500]$ GHz and $L_2=12, 14, 16$ $\mu$m.
Suggested by Fig. \ref{fig3}(d, e), we select $L_2=14$ $\mu$m for $n_\text{r,rf}=n_\text{g,opt}=2.25$ and $Z_\text{c}=50$ $\Omega$ at $\omega_\text{rf}/2\pi=400$ GHz.
Figure \ref{fig3}(f) shows the impact of electrode materials on $\alpha_\text{rf}$, i.e.  gold (Au) with tan$\delta$, Cu with tan$\delta$, and Cu without tan$\delta$.
As Cu ($\sigma_\text{Cu}\approx 5.8\times10^{7}$ S/m) has higher conductivity than Au ($\sigma_\text{Au}\approx 4.5\times10^{7}$ S/m), Cu electrodes enable lower RF attenuation (smaller $\alpha_\text{rf}$).
Meanwhile, tan$\delta$ of SiO$_2$ and LiNbO$_3$ increases $\alpha_\text{rf}$ for $\omega_\text{rf}/2\pi>100$ GHz.
The electric ($E_\text{x}$) and magnetic ($H_\text{x}$ and $H_\text{y}$) fields of Si and LiNbO$_3$ TW-MZMs are found in Supplementary Note 4. 

We also simulate the electrode design in Ref.~\cite{ZhangY:24} and obtain $n_\text{r,rf}=2.26$, $\alpha_\text{rf}=13$ dB/cm and $Z_\text{c}=46$ $\Omega$ with $\omega_\text{rf}/2\pi=300$ GHz.
These values closely match the experimental result of $n_\text{r,rf}=2.26$, $\alpha_\text{rf}=14$ dB/cm and $Z_\text{c}=46$ $\Omega$ in Ref.~\cite{ZhangY:24}.
Further validations are provided in Supplementary Note 8.
Compared to the simulation of 1000-$\mu$m-long electrodes in Ref.~\cite{ZhangY:24}, here we only simulate half of a 20-$\mu$m-long unit cell. 
Our simulation saves substantial computational resources and is more than 100 times faster.

%%%%%%%%%%%%%%%%%%%%%%%%%%%%%%%%%%%%%%%%%%%%%%%%%%%%%%%%%%%%%%%%%%%%%%%%%%%%%%%
\section{Optoelectronic co-simulation to solve TCME}
%%%%%%%%%%%%%%%%%%%%%%%%%%%%%%%%%%%%%
\begin{figure*}[t!]
\centering
\includegraphics[width=15.5cm]{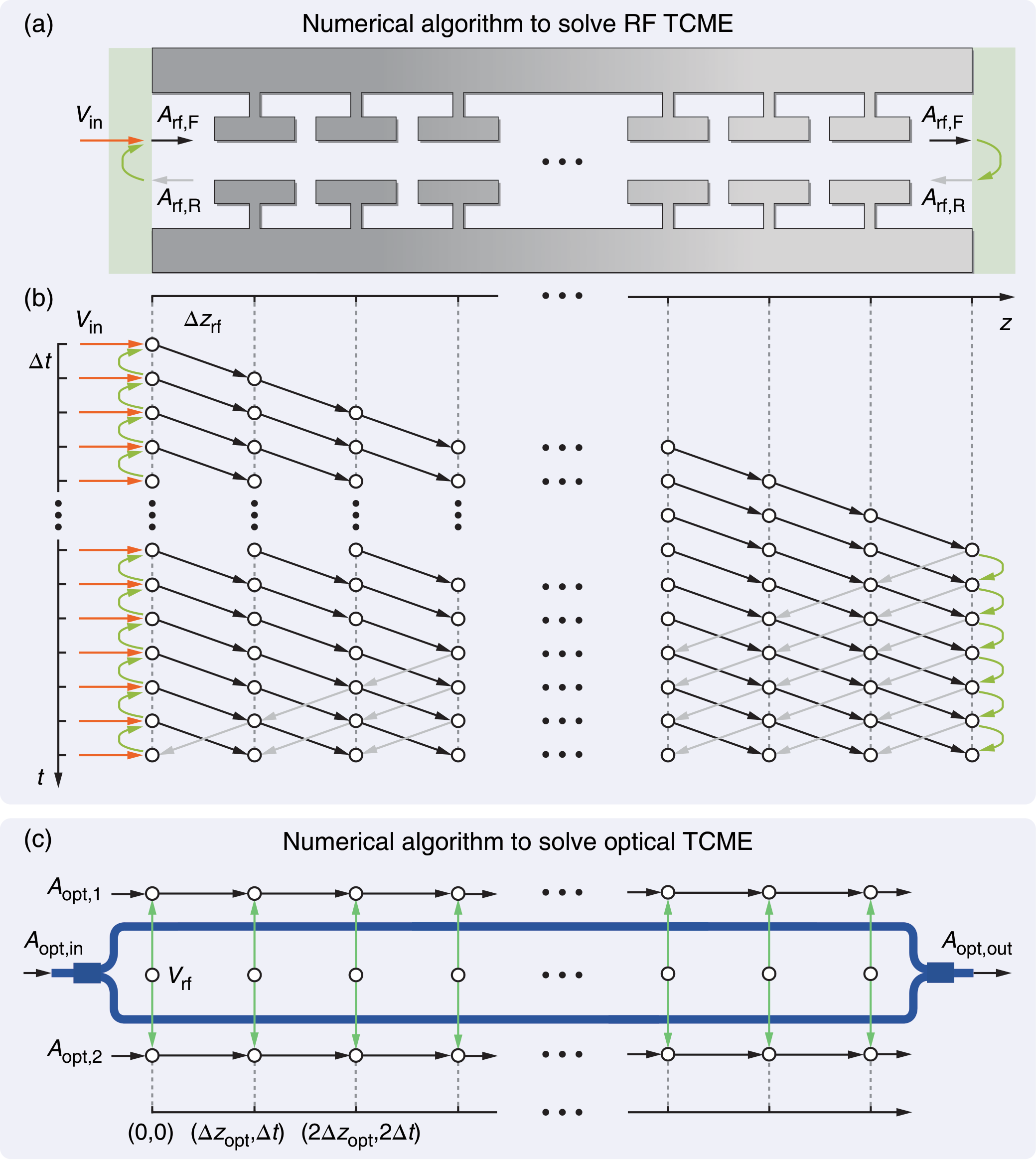}
\caption{Optoelectronic co-simulation to solve TCME.
(a) Numerical algorithm to solve the RF TCME.
Green regions mark the boundary condition for $A_\text{rf,F}$ and $A_\text{rf,R}$.
(b) 2D time-space mesh grid to solve the RF TCME. 
Each circle represents a node storing the values of $A_\text{rf,F}$ and $A_\text{rf,R}$. 
Black and gray arrows mark the sequence to calculate $A_\text{rf,F}$ and $A_\text{rf,R}$. 
Green arrows mark the boundary condition. 
Orange arrows mark the input voltage signal $V_\text{in}$.   
(c) Numerical algorithm to solve the optical TCME.
The input optical wave $A_\text{opt,in}$ is divided into $A_\text{opt,1}$ and $A_\text{opt,2}$. 
Black arrows mark the sequence to calculate $A_\text{opt,1}$ and $A_\text{opt,2}$. 
Here $V_\text{rf}$ to represent $V_\text{rf,F}$, $V_\text{rf,R}$ and $V_\text{dc}$, and cyan arrows represent nonlinearity terms of $V_\text{rf}$.
}
\label{fig4}
\end{figure*}
%%%%%%%%%%%%%%%%%%%%%%%%%%%%%%%%%%%%%%%

With the acquired coefficients via eigenmode solvers, we next numerically solve the TCME Eq. (\ref{Eq.RFF_TCME}--\ref{Eq.Opt_TCME}) using optoelectronic co-simulation. 
Figure \ref{fig4} delineates the numerical algorithms to calculate $A_\text{rf,F}$, $A_\text{rf,R}$ and $A_\text{opt}$. 
Figure \ref{fig4}(a) shows the periodic T-shaped RF electrodes as an RF waveguide. 
The green regions mark the boundary condition (see Supplementary Note 5). 
We use two time-space mesh grids of identical temporal grid size $\Delta t=0.5$ ps, sufficiently small to achieve numerical convergence.
Thus, the spatial grid sizes are different given by the different wave speeds, i.e. $\Delta z_\text{rf}=v_\text{g,rf}\Delta t$ and $\Delta z_\text{opt}=v_\text{g,opt}\Delta t$.
Figure \ref{fig4}(b) depicts the 2D mesh grid used to solve the RF TCME Eq. (\ref{Eq.RFF_TCME}, \ref{Eq.RFR_TCME}). 
Each circle represents a node storing the values of $A_\text{rf,F}$ and $A_\text{rf,R}$, whose initial values are zero.
Black and gray arrows mark the calculation sequence. 
Green arrows mark the boundary condition. 
Orange arrows mark the input voltage signal $V_\text{in}$.   

The algorithm starts with the calculation of $A_\text{rf,F}$, which is stimulated by $V_\text{in}$ via the left boundary condition and follows the black arrows. 
The calculation of $A_\text{rf,R}$ is triggered by $A_\text{rf,F}$ on the right boundary condition due to impedance mismatch.
Ultimately, the algorithm yields the distribution of $V_\text{rf,F}$ and $V_\text{rf,R}$ on the mesh grid based on Eq. (\ref{V_def}).

Figure \ref{fig4}(c) shows that the input optical wave $A_\text{opt,in}$ is divided into $A_\text{opt,1}$ and $A_\text{opt,2}$ in the MZI's two arms.
Black arrows mark the sequence to calculate $A_\text{opt,1}$ and $A_\text{opt,2}$. 
Cyan arrows represent Eq. (\ref{Eq.Popt}).
For brevity, we use $V_\text{rf}$ to represent $V_\text{rf,F}$, $V_\text{rf,R}$ and $V_\text{dc}$.
Since $\Delta z_\text{opt}\neq\Delta z_\text{rf}$, we use interpolation to re-sample the distribution of $V_\text{rf}$ along the $z$-direction,
consistent with the 2D mesh grid used to calculate $A_\text{opt,1}$ and $A_\text{opt,2}$.
As $\omega_\text{opt}\gg\omega_\text{rf}$, instead of using a finer grid size $\Delta z_\text{opt}$, we use an equivalent approach to transform the differential form of $A_\text{opt,1}$ and $A_\text{opt,2}$ into the exponential differential form.
The output optical wave is $A_\text{opt,out}$ recombining $A_\text{opt,1}$ and $A_\text{opt,2}$.
The exact differential forms of $A_\text{rf,F}$, $A_\text{rf,R}$, $A_\text{opt,1}$ and $A_\text{opt,2}$ are found in Supplementary Note 6.

%%%%%%%%%%%%%%%%%%%%%%%%%%%%%%%%%%%%%%%%%%%%%%%%%%%%%%%%%%%%%%%%%%%%%%%%%%%%%%%
\section{Simulation and experimental validation of eye diagrams}

%%%%%%%%%%%%%%%%%%%%%%%%%%%%%%%%%%%%%%
\begin{figure*}[t!]
\centering
\includegraphics[width=15.5cm]{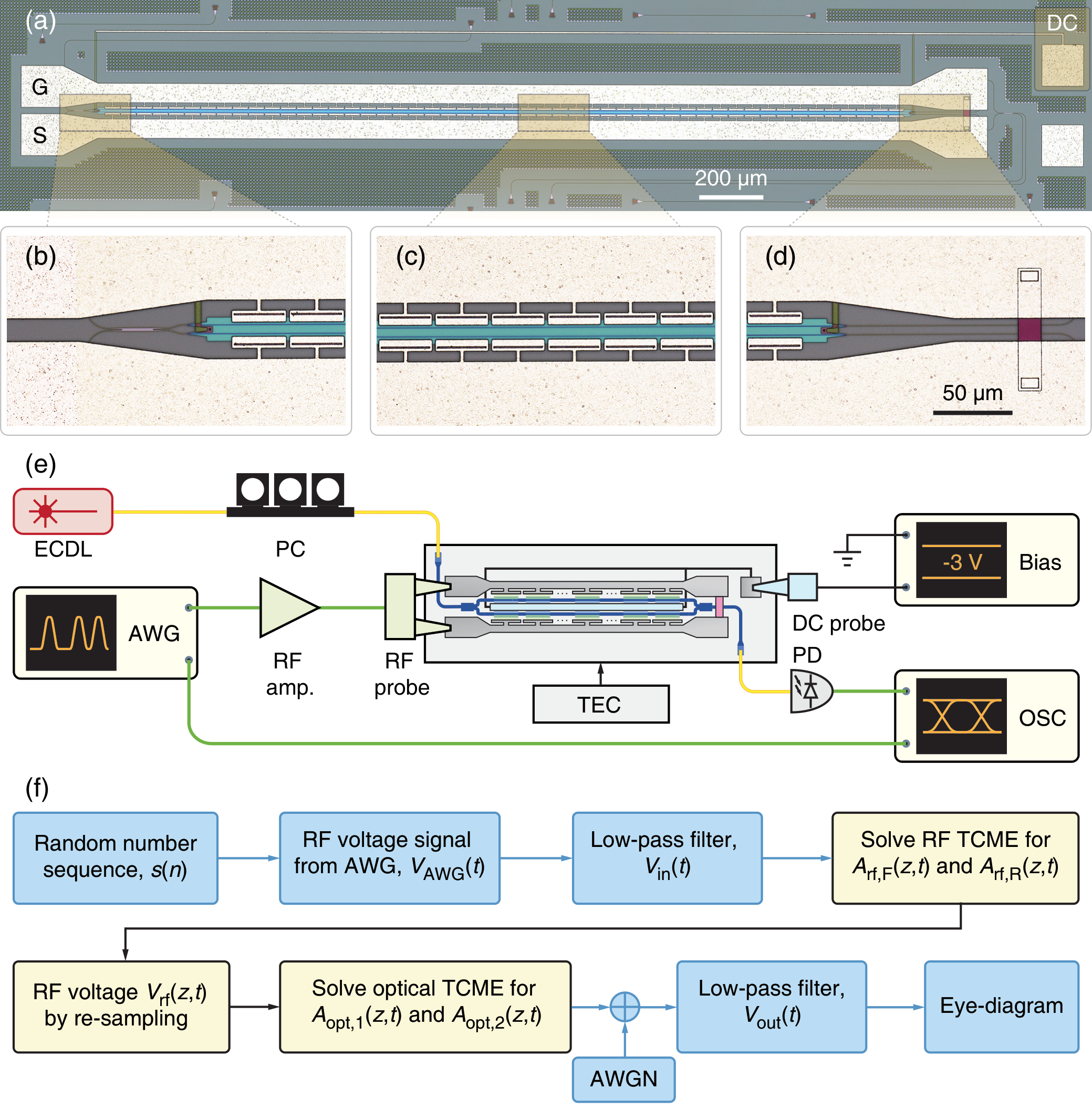} 
\caption{
Simulation and experimental validation of eye diagrams.
(a) Optical microscope images of a Si TW-MZM.
(b-d) Zoom-in views of the Si TW-MZM, highlighting the MMI (b),  the electrode's unit cell (c), and the matched resistor (d).
(e) Experimental setup to measure eye diagrams of the Si TW-MZM.
(f) Simulation flowchart outlining the process for generating eye diagrams.
}
\label{fig5}
\end{figure*}
%%%%%%%%%%%%%%%%%%%%%%%%%%%%%%%%%%%%%%

%%%%%%%%%%%%%%%%%%%%%%%%%%%%%%%%%%%
\begin{figure*}[t!]
\centering
\includegraphics[width=15.5cm]{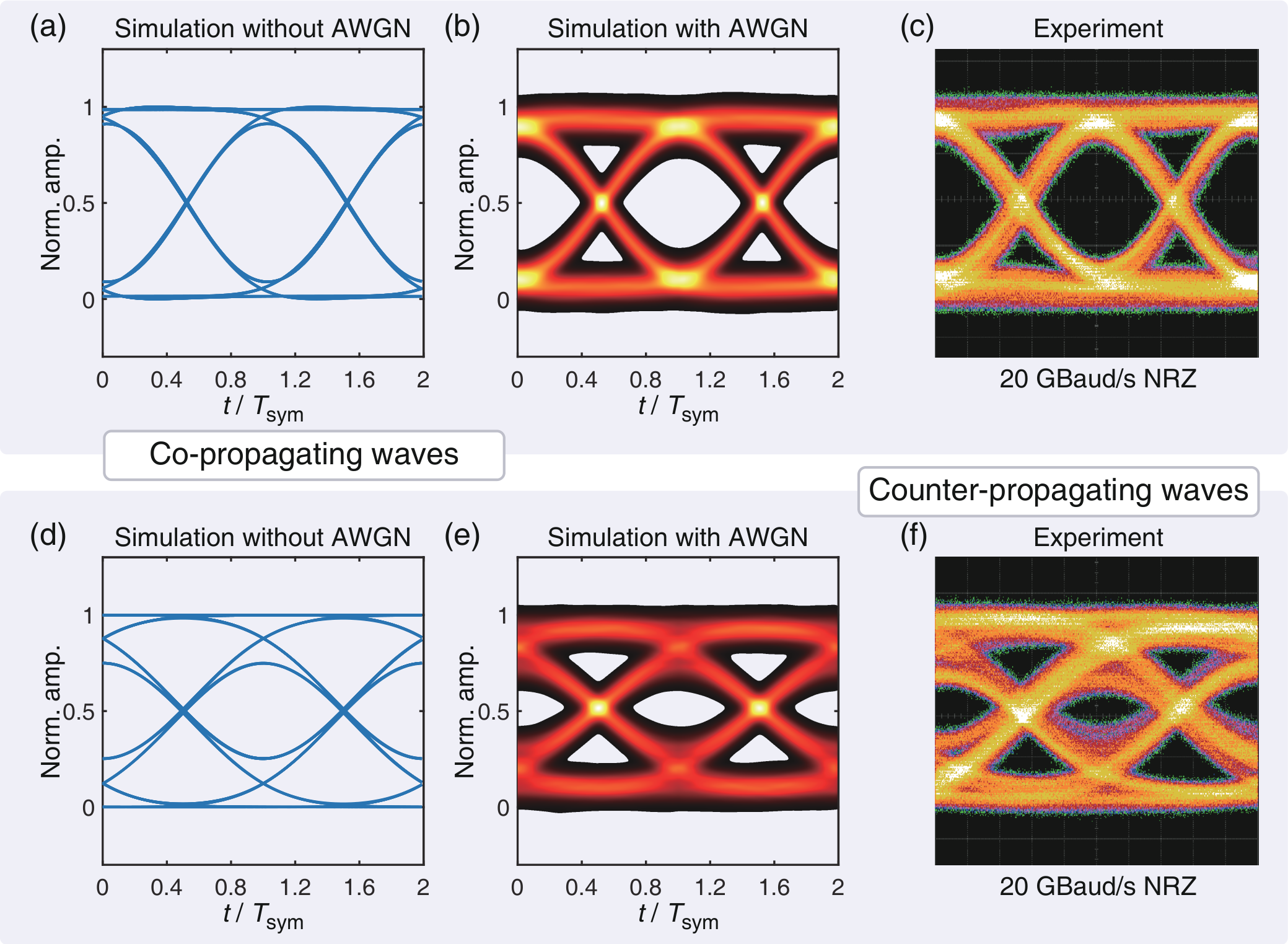} 
\caption{
Simulated and experimental eye diagrams.
With co-propagating waves, the simulated $\text{ER}=6.5$ dB and $Q=7.5$ in (b) agree with the measured $\text{ER}=6.4$ dB and $Q=7.3$ in (c).
With counter-propagating waves, the simulated $\text{ER}=4.8$ dB and $Q=3.3$ in (e) also agree with the measured $\text{ER}=4.4$ dB and $Q=3.2$ in (f).
}
\label{fig6}
\end{figure*}
%%%%%%%%%%%%%%%%%%%%%%%%%%%%%%%%%%%

Finally, the TW-MZM's eye diagram can be obtained with the solved TCME. 
In this section, we first elaborate the experimental measurement of a Si TW-MZM, and then show simulation results that are validated by the experiment. 
Figure \ref{fig5}(a) shows an optical microscope image of a 2.5-mm-long Si TW-MZM fabricated using a commercial CMOS foundry process.
Figure \ref{fig5}(c) highlights the RF electrodes with PN junctions. 
Figure \ref{fig5}(b) shows the upper and lower electrodes connecting to a GS pad for RF input.  
Figure \ref{fig5}(d) shows the matched resistor for RF absorption. 
A length difference of 100 $\mu$m between the MZI's two arms enables  3-dB operating point tuning by varying the laser frequency.
As shown in Fig. \ref{fig5}(b,d), the DC electrical circuit connects to the P++ region and provides reverse-biased voltage for the PN junctions.

Figure \ref{fig5}(e) displays the experimental setup for eye-diagram measurement in digital communication using our Si TW-MZM. 
An arbitrary waveform generator (AWG) serves as the RF source providing non-return-to-zero (NRZ) modulation signals. 
After RF amplification and transmission in RF cables and probes, the RF peak-to-peak voltage is estimated as $V_\text{pp}=5$ V . 
The AWG is also connected to and synchronizes an oscilloscope (OSC). 
A thermo-electric cooler (TEC) maintains the TW-MZM at $25^\circ$C. 
The DC source supplies $V_\text{dc}=-3$V bias, with one terminal to ground. 
The output wavelength of the external-cavity diode laser (ECDL) is set at the TW-MZM's operating point of 1549.5 nm. 
A polarization controller (PC) is used to optimize the input optical signal's polarization. 
A photodetector (PD) collects the modulated light from the TW-MZM. 
The eye diagram is displayed by the OSC and shown in Fig. \ref{fig6}(c, f). 

In parallel, we simulate the eye diagram based on the experimental setup. 
The simulation flow chart is illustrated in Fig. \ref{fig5}(f). 
We generate NRZ modulation signals from a sequence of random numbers $s(n)$.
The RF voltage signal $V_\text{AWG}(t)$ is created based on $s(n)$.
Considering the actual bandwidth of our AWG and RF amplifier, a digital low-pass filter is used to limit the bandwidth of $V_\text{AWG}(t)$ to below 20 GHz. 
The filtered result is $V_\text{in}(t)$. 
With the previously acquired numerical solution of TCME,  the values of  $A_\text{rf,F}$, $A_\text{rf,R}$, $A_\text{opt,1}$ and $A_\text{opt,2}$ are input to the eye-diagram simulation.
Additive white Gaussian noise (AWGN), characterized by the signal-to-noise ratio (SNR, described latter) \cite{HaoX:14}, is introduced to mimic experimental noise.  
The signal with AWGN passes through a low-pass filter mimicking the RF bandwidth below 20 GHz of the PD and OSC.
The filtered signal is visualized as an eye diagram by accumulating data within two digital symbol periods $T_\text{sym}$.

Figure \ref{fig6} compares the simulated and experimental eye diagrams, for two configurations: 
the co-propagating optical and RF waves in Fig. \ref{fig6}(a--c), and the counter-propagating waves in Fig. \ref{fig6}(d--f).
This bidirectional propagation scheme is widely used in designing Michelson-interferometer modulators \cite{XuM:19}, modulator-based gyroscopes \cite{PogorelayaD:19}, and isolators \cite{YuM:23}.
Figure \ref{fig6}(a,d,b,e) shows the simulated eye diagrams without or with AWGN. 
For counter-propagation waves, the sequence to calculate $A_\text{opt,1}$ and $A_\text{opt,2}$ is reversed as shown in Fig. \ref{fig4}(c). 
Counter-propagating waves are experimentally realized by interchanging the input (ECDL and PC) with the output (PD and OSC) in Fig. \ref{fig5}(e). 
Figure \ref{fig6} evidences that the eye diagram with counter-propagating waves is worse. 

In the simulation of co-propagating waves, we set the SNR of AWGN to 22.5 dB. 
The resulting eye diagram, shown in Fig. \ref{fig6}(b), exhibits 6.5 dB extinction ratio (ER) and a quality factor of $Q=7.5$. 
These values agree with the experimental result of 6.4 dB ER and $Q=7.3$ in Fig. \ref{fig6}(c).
Applying the same SNR (22.5 dB) to counter-propagating waves, the simulated eye diagram in Fig. \ref{fig6}(e) shows a 4.8 dB ER and $Q=3.3$. 
Again, these values agree with the experimental result of 4.4 dB ER and $Q=3.2$ in Fig. \ref{fig6}(f). 

%%%%%%%%%%%%%%%%%%%%%%%%%%%%%%%%%%%%%%%%%%%%%%%%%%%%%%%%%%%%%%%%%%%%%%%%%%%%%%%
\section{Conclusion}
In conclusion, we present a universal perspective to understand and analyze high-speed TW-MZMs using nonlinear optics and complex band structures. 
Under this perspective, we design, simulate and experimentally validate high-speed TW-MZMs based on Si and LiNbO$_3$. 
Our approach shows unambiguous advantages in simplicity, accuracy and efficiency over conventional methods. 
Detailed comparisons highlighting the key advantages of our method over previous approaches are summarized in Supplementary Note 7.
Though showcased on Si and LiNbO$_3$ TW-MZMs, the proposed method is highly adaptable and can essentially be applied to nearly any integrated material platform.
In addition, by considering the imaginary part of nonlinear coefficients, our approach is equally practical for TW-MZMs based on electro-absorption materials such as graphene \cite{LiuM:11}, SiGe \cite{LiuJ:08} and III-V \cite{HirakiT:21}. 
For integrated semiconductor platforms such as Si, Ge and III-V, this approach based on the equivalent nonlinearity of PN junctions in both the RF and optical domains can be instrumental for nonlinear and quantum microwave photonics \cite{MarpaungD:14,SahuR:23}.
Finally, our optoelectronic co-simulation establishes a wave-based framework, facilitating the synergy and convergence of electronics and photonics. 
With electrode designs capable of reaching frequencies above 500 GHz, our approach offers a viable route to constructing future high-speed TW-MZMs for millimeter-wave and terahertz applications.

%%%%%%%%%%%%%%%%%%%%%%%%%%%%%%%%%%%%%%%%%%%%%%%%%%%%%%%%%%%%%%%%%%%%%%%%%%%%%%%%%%%%
\vspace{0.3cm}
\begin{acknowledgments}
The authors thank Lan Gao, Xue Bai, Baoqi Shi, Jinbao Long, and Yihan Luo for their suggestions on the manuscript.
This work was supported by the National Key R$\&$D Program of China (Grants No. 2024YFA1409300), National Natural Science Foundation of China (No. 12261131503, 62405202), Innovation Program for Quantum Science and Technology (2023ZD0301500), Shenzhen Science and Technology Program (No. RCJC20231211090042078), Shenzhen-Hong Kong Cooperation Zone for Technology and Innovation (HZQB-KCZYB2020050), Guangdong-Hong Kong Technology Cooperation Funding Scheme (No. 2024A0505040008).
Data underlying the results presented in this paper are not publicly available at this time but may be obtained from the authors upon reasonable request.
\end{acknowledgments}

%%%%%%%%%%%%%%%%%%%%%%% References %%%%%%%%%%%%%%%%%%%%%%%%%
%%%%%%%%%% If using BibTeX:
\bibliography{MZM_paper_ref}
\clearpage
%%%%%%%%%%%%%%%%%%%%%%%%%%%%%%%%%%%%%%%%%%%%%%%%%%%%%%%%%%%%%%%%%%%%%%%%%%%%%%%%%%%%%%%%%%%%%%%%
\end{document}